\documentclass[a4paper,fleqn,usenatbib]{mnras}

\usepackage{newtxtext,newtxmath}

\usepackage[T1]{fontenc}
\usepackage{ae,aecompl}

\usepackage{graphicx}	
\usepackage{amsmath}	
\usepackage{amssymb}	

\defcitealias{Corral-Santana2013}{CS13}
\defcitealias{MataSanchez2015}{MS15}
\usepackage{multirow}
\usepackage{threeparttable}

\title[Dip-resolved spectroscopy of Swift~J1357.2-0933]{An equatorial outflow in the black hole optical dipper Swift~J1357.2-0933}

\author[F. Jim\'{e}nez-Ibarra et al.]{F. Jim\'{e}nez-Ibarra,$^{1,2}$\thanks{E-mail: felipeji@iac.es}
T. Mu\~{n}oz-Darias,$^{1,2}$
J. Casares,$^{1,2}$
M. Armas Padilla,$^{1,2}$ and\newauthor
J. M.  Corral-Santana$^{3}$ 
\\
$^{1}$Instituto de Astrof\'{i}sica de Canarias, V\'{i}a L\'{a}ctea, La Laguna, E-38205, Santa Cruz de Tenerife, Spain\\
$^{2}$Departamento de Astrof\'{i}sica, Universidad de La Laguna, E-38206, Santa Cruz de Tenerife, Spain\\
$^{3}$European Southern Observatory (ESO), Alonso de C\'{o}rdova 3107, Vitacura, Casilla 19, Santiago, Chile\\}

\date{Accepted XXX. Received YYY; in original form ZZZ}

\pubyear{2019}

\begin{document}
\label{firstpage}
\pagerange{\pageref{firstpage}--\pageref{lastpage}}
\maketitle

\begin{abstract} 
We present high-time resolution optical spectroscopy and imaging of the black hole transient Swift~J1357.2-0933 during its 2017 outburst. The light curves show recurrent dips resembling those discovered during the 2011 outburst. The dip properties (e.g duration and depth) as well as the evolution of their recurrence time are similar to those seen in 2011. Spectra obtained during the dips are characterised by broad and blue-shifted absorptions in Balmer and \ion{He}{ii}. The absorptions show core velocities of $\sim-800$~km~s$^{-1}$ and terminal velocities approaching $\sim$3000~km~s$^{-1}$ i.e. in the upper-end of wind velocities measured in other black hole transients (both at optical and X-ray wavelengths). Our observations suggest that the dips are formed in a dense and clumpy outflow, produced near the disc equatorial plane and seen at high inclination. We also study the colour evolution and observe that, as it has been previously reported, the source turns bluer during dips. We show that this is due to a gradual change in the slope of the optical continuum and discuss possible implications of this behaviour.
\end{abstract}



\section{Introduction}
Accretion is the governing mechanism in a large number of astrophysical contexts from planetary formation to active galactic nuclei. In the case of black hole X-ray binaries (i.e interacting binaries where a companion star transfer matter onto a black hole) accretion results in the most efficient mechanism of converting matter into energy. In these systems accretion proceeds via an accretion disc. The high temperature reached in this disc ($\sim$10$^{7}$~K) makes them luminous X-ray sources, that also radiate at lower energies, mainly due to thermal reprocessing of the X-ray photons in different binary regions. Accretion in X-ray binaries exhibit timing properties accessible to human time-scales, showing variability ranging from sub-seconds to months \citep[see][for reviews]{Remillard2006,vanderKlis2006,Belloni2011}.

Among black hole X-ray binaries, transient systems (BHTs), alternate long periods of faint quiescence with bright and short (weeks to months) outbursts. These episodes are triggered by a sudden increase of mass accretion onto the black hole and reach luminosities typically above $\sim$10 per cent of the Eddington limit. A wide variety of outflowing  phenomenology is also common to BHTs. This consists of collimated radio jets and both, low and highly ionised winds. In general, the properties of the outflows have been found to be strongly coupled to those of the accretion flow \citep[see e.g.,][]{Gallo2003,Corbel2003,Ponti2016,Fender2016,Munoz-Darias2019}. 

Swift~J1357.2-0933 is a BHT  discovered during an outburst episode in 2011 \citep{Krimm2011,ArmasPadilla2013a}. Its orbital period is among the shortest of its class ($\sim$2.8~h) and it exhibits the broadest disc emission lines among BHTs, a strong indication for a high binary inclination \citep[][hereafter \citetalias{Corral-Santana2013}]{Corral-Santana2013}. A  black hole mass $>$ 9.3 $M_{\odot}$ has been inferred by applying empirical scaling relations to the full-width at half-maximum (FWHM) of the H$\alpha$ line \citep[][hereafter \citetalias{MataSanchez2015}; \citealt{Casares2016}]{MataSanchez2015}.

During the decay of its 2011 outburst \citetalias{Corral-Santana2013} observed regular dips in the optical light curve. The dips were profound (up to $\sim$0.8~mag depth, with characteristic durations of $\sim$2~min) and recurred with increasing quasi-periodicity as the outburst declined. These features were interpreted as caused by obscuration of inner disc regions by a vertical structure that moved outwards as the outburst proceeded \citepalias{Corral-Santana2013,MataSanchez2015}. The nature of the obscuring structure remains unclear and has been the subject of much debate \citep{ArmasPadilla2014a,Torres2015,Beri2019}.  

In 2017, the system went into another outburst \citep{Drake2017} and new optical dips, evolving in a similar way as those reported in 2011, were observed again \citep{Paice2019}. In order to investigate the puzzling nature of the dips we used the 10.4m Gran Telescopio Canarias (GTC) and obtained high time resolution spectroscopy through the dips for the first time.

\section{Observations and data reduction} \label{sec:obsred}

\subsection{Optical spectroscopy}
We carried out optical spectroscopy of Swift~J1357.2-0933 over 3 nights during the 2017 outburst, using the Optical System for Imaging and low-Intermediate Resolution Integrated Spectroscopy \citep[OSIRIS;][]{Cepa2000} mounted on the 10.4m GTC at the Observatorio del Roque de los Muchachos (ORM; La Palma, Spain). On 4 May 2017, we obtained 2 exposures of 400~s each using the R1000B optical grism in the spectral range 3630--7500~\AA. Combined with a slit width of 1 arcsec this yields a velocity resolution of $\sim350$~km s$^{-1}$, as measured from the FWHM of the \ion{O}{i} airglow line at $\sim5577$~\AA. In addition, we obtained fast timing spectroscopy with the R300B grism on the nights of 12 and 16 June 2017 in order to resolve the optical dips. For this purpose, we used individual exposure times of $\sim$13~s which, together with overheads, resulted in time resolutions in the range of $22-37$~s. We collected a total of 304 spectra (165 on 12 June and 139 on 16 June), covering the spectral range 3600--7200~\AA. We employed two slit widths of 1 and 0.6 arcsec for the nights of 12 and 16 June. This set-up provided spectral resolutions ranging from 590 to 970~km s$^{-1}$ (determined using sky lines). We note that the seeing was in the range of $\sim 1$--$1.5$~arcsec and $\sim 0.7$--$1.3$~arcsec during the first and second epoch, respectively, and therefore the spectral resolution was always limited by the slit-width. We applied bias and flatfielding corrections using \textsc{iraf} standard routines, while cosmic rays were removed with L.A.Cosmic \citep{vanDokkum2001}. The pixel-to-wavelength calibration was handled using arc lamp exposures taken on each observing night (HgAr+Ne lamps for the R1000B grism and HgAr+Ne+Xe for the R300B). All the spectra were corrected for velocity drifts introduced by instrumental flexure ($<$ 110~km~s$^{-1}$) using \textsc{molly} software and the \ion{O}{i} $\sim5577$ \AA\ sky emission line.

The slit was oriented to include a nearby star, which was used to monitor and correct for possible slit losses caused by variable seeing conditions. This was done through scaling the spectrum of Swift~J1357.2-0933 to that of the on-slit star in each spectrum. In order to obtain absolute fluxes, we calibrated the best on-slit star spectrum of the series (i.e. the one obtained under the best conditions) against the flux standard GD 153 \citep{Bohlin1995}, which was observed the same night but using a wider slit (2.52 arcsec). To account for the different slit-width between the on-slit star and GD 153, we refine the flux calibration of the former by using Sloan-g band photometry obtained from the acquisition image. This was achieved through synthetic photometry obtained by convolving the spectrum of the on-slit star with the Sloan-g filter bandpass. Finally, the calibrated spectrum of the on-slit star was used to extend the absolute flux calibration to every scaled spectra of Swift~J1357.2-0933. We note that even if the absolute fluxes values were affected by a small systematic offset, the relative fluxes between the different spectra should be highly accurate.

\subsection{Optical photometry}
We also used the Rapid Imager for Surveys of Exoplanets (RISE) fast-readout camera attached to the 2m Liverpool Telescope (LT, at the ORM) to obtain high time-resolution photometry of Swift~J1357.2-0933. Observations were performed on 7 different nights between 15 May and 14 July 2017, using the OG515 and KG3 filters ($\sim$ V+R band). The light curve on the night of 12 June 2017 is strictly simultaneous with the GTC  spectroscopy. The fast readout time of RISE ($\sim0.04$~s) minimised the overhead between exposures, resulting in time resolutions near the exposure time ($\sim5$~s). The data reduction was completed using the data reduction pipeline for the optical imaging component of the Infrared-Optical suite\footnote{http://telescope.livjm.ac.uk/TelInst/Pipelines/\#ioo}. Flux calibration was performed against field stars catalogued in PanSTARRS and using \textsc{astropy-photutils} based routines \citep{Bradley2019} in the PanSTARRS-1 broad-band filter \textit{r}.

\section{Analysis} \label{sec:analy}

\subsection{Optical dips} \label{subsec:opdibs}
\begin{figure*}
\includegraphics[width=\textwidth]{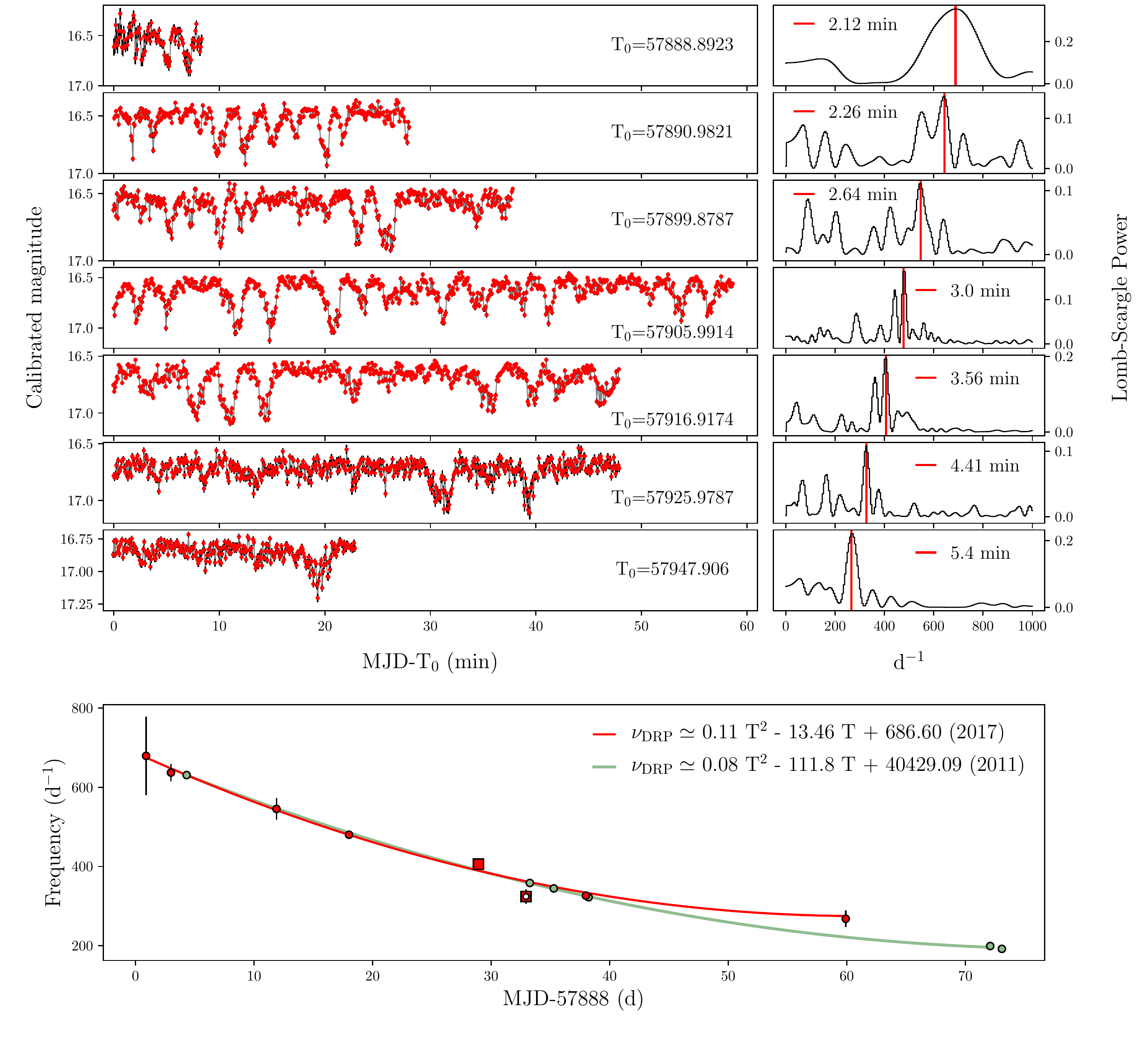}
\caption{Top panels: optical light curves of Swift~J1357.2-0933 (left) and their corresponding power density spectra obtained from the Lomb-Scargle analysis (right). The frequency of the highest peak is indicated by a red vertical line. Bottom panel: time evolution of the DRP frequency. The solid lines indicate the best parabolic fits to the data points. Our data is indicated in red, while data from \citetalias{Corral-Santana2013} is shown in green. The 2011 curve has been shifted in time by matching the first point of the series to the interpolation of the 2017 parabolic fit. Red squares indicate the frequencies obtained from the GTC continuum light curves. The empty square indicates the frequency concurrently measured from the LT data and the GTC continuum light curve.}
\label{fig:phot}
\end{figure*}

Fig. \ref{fig:phot} (top-left panels) presents the seven LT light curves. We observe numerous dip events lasting typically $\sim 2$~min, similar to those witnessed in the 2011 outburst by \citetalias{Corral-Santana2013}. The dips are up to $\sim0.5$ mag deep. In addition, we produced two light curves (one per epoch) from the continuum of the high-time resolution GTC spectra. To this end, we integrated the flux density within 2 apertures defined in featureless spectral regions (see Fig. \ref{fig:spect}, top panel). The GTC continuum light curves are shown in the middle panel of Fig. \ref{fig:spect}.

In order to explore the timing properties of the dips we produced Lomb-Scargle periodograms of the 9 light curves using the \textsc{Lomb-Scargle-python} class \citep{VanderPlas2018}. The periodograms of the seven LT epochs are shown in the right panel of Fig. \ref{fig:phot}. The highest peak reveals a dip recurrence period (DRP) that increases from 2.09 to 5.41 min over 59 days (see Table \ref{tab:DRP}). Note that the DRP concurrently measured from the LT and GTC light curves on 12 June 2017 are consistent within errors.

The evolution of the frequency (from DRP) as a function of time, together with the best parabolic fit, are shown in red in Fig. \ref{fig:phot} (bottom panel). In the same panel, green points represent the data and best fit from \citetalias{Corral-Santana2013} (2011 outburst). In order to compare the evolution of the two frequency tracks, a time shift was applied to the 2011 data. This was obtained by matching the frequency of the first 2011 point to the interpolation of the 2017 frequency curve. A visual comparison reveals a remarkable analogy in both evolutions. The similarity in both dips properties and DRP evolution strongly suggests that the phenomenon producing the dips in 2017 is the same than in 2011.


\begin{table}
	\centering
	\caption{Time evolution of the DRP}
\begin{threeparttable}
		\begin{tabular}{lccc} 
		\hline
		Date & MJD & \multicolumn{2}{c}{DRP} \\	
						 & (days) &( cycle d$^{-1}$)& (min)\\				
		\hline
15-05-2017 & 57888.8952156 & 679 $\pm$ 99 & 2.12 $\pm$ 0.31 \\
17-05-2017 & 57890.9918298 & 637 $\pm$ 22 & 2.26 $\pm$ 0.08 \\
26-05-2017 & 57899.8918105 & 545 $\pm$ 27 & 2.64 $\pm$ 0.13 \\
02-06-2017 & 57906.0117608 & 480 $\pm$ 8 & 3.00 $\pm$ 0.05 \\
12-06-2017 & 57916.9340031* & 404 $\pm$ 11 & 3.56 $\pm$ 0.10 \\
12-06-2017 & 57916.9376235 & 406 $\pm$ 9 & 3.55 $\pm$ 0.08 \\
16-06-2017 & 57920.9410283* & 324 $\pm$ 18 & 4.44 $\pm$ 0.25 \\
21-06-2017 & 57925.9953267 & 326 $\pm$ 11 & 4.41 $\pm$ 0.15 \\
13-07-2017 & 57947.913911 & 268 $\pm$ 21 & 5.4 $\pm$ 0.4 \\
		\hline
\end{tabular}
\begin{tablenotes}
\item[]*Obtained from the GTC continuum light curve
\end{tablenotes}
\end{threeparttable}
\label{tab:DRP}
\end{table}


\subsection{Spectral analysis}\label{sec:spec}
\begin{figure*}
\includegraphics[width=\textwidth]{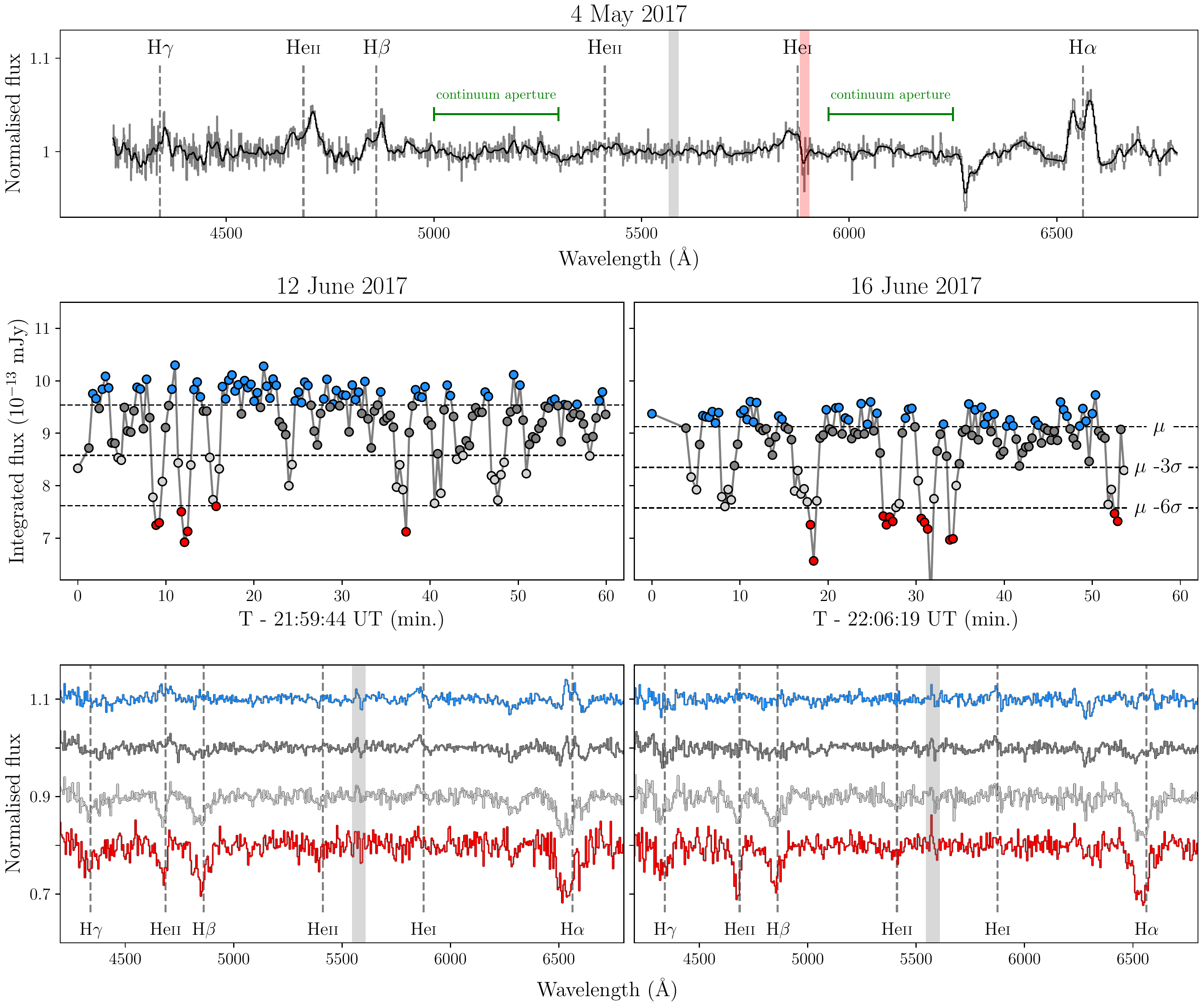}
\caption{Top panel: continuum normalised spectrum of Swift~J1357.2-0933 obtained on 4 May 2017. A Gaussian smoothed version is over-plotted in black. The identified emission lines are indicated by dashed lines. The red shaded line indicates the \ion{Na}{i}-doublet at 5890-5896 \AA. The continuum regions defined for the light curve extraction are indicated by green segments. Middle panel: continuum light curves extracted from the high-time resolution spectra. Coloured dots indicate non-dip spectra (blue) and 3 levels of dip spectra (dark gray, light gray, and red). The horizontal dashed lines mark the limits defining each group as described in the text. Bottom panels: normalised dip, and non-dip spectra following the light curve colour code defined above. A vertical offsets of 0.1 has been added to the normalised spectra for clarity. The $\sim$5577 \AA\ sky emission line region is indicated in all the spectra as a gray thick line.}
\label{fig:spect}
\end{figure*}

In Fig. \ref{fig:spect} (top panel) we present the averaged spectrum of Swift~J1357.2-0933 obtained on 4 May 2017 with the R1000B grism. We identify broad emission lines corresponding to the Balmer series (up to H$\gamma$), \ion{He}{i}-5876 \AA, and \ion{He}{ii} (4686 \AA\ and 5411 \AA). All the lines display double-peaked profiles, with the exception of \ion{He}{i}-5876 \AA, where the red edge is affected by the \ion{Na}{i}-doublet interstellar absorption at 5890--5896 \AA. The emission lines are all remarkably broad (H$\alpha$ FWHM $\sim$ 3000~km~s$^{-1}$), similarly to what was seen during the 2011 outburst \citepalias[H$\alpha$ FWHM $\sim$ 3300~km~s$^{-1}$,][]{Corral-Santana2013}. We determined the centroid velocity of the H$\alpha$ line by fitting a two-Gaussian model (to account for the double-peaked profile) over the R1000B normalised spectrum. We obtained $-130\pm17$~km~s$^{-1}$, which is consistent with the systemic velocity ($\gamma$) measured in the previous outburst \citepalias[$\gamma \sim -150$~km~s$^{-1}$,][]{Corral-Santana2013}. On the other hand, we obtain different velocities from the R300B spectra, $\sim  -300$~km~s$^{-1}$ on 12 June and 0~km~s$^{-1}$ on 16 June. Variations in the centroid velocity of the H$\alpha$ line have been observed before and interpreted as signatures of a precessing accretion disc \citepalias{MataSanchez2015}.

Based on the continuum light curve (see Sec. \ref{subsec:opdibs}), we define two groups of spectra that we call {\it dip} and {\it non-dip} spectra hereafter. In order to select them we have computed the mean and the standard deviation ($\sigma$) of the continuum light curve using a sigma clipping algorithm \citep[\textsc{astropy.stats}][]{Astropy2013}. Points above the mean are considered non-dip spectra, while points below the mean level are labelled as dip spectra. We have also divided dip spectra into 3 further groups separated by the $3\sigma$ and the $6\sigma$ levels (see Fig. \ref{fig:spect}, middle panel). We subsequently combined the selected spectra to produce four averaged spectra per night (three dip and one non-dip). The resulting spectra are shown in Fig. \ref{fig:spect} (bottom panel). 

The non-dip averaged spectra show double-peaked H$\alpha$ profiles. Conversely, the H$\beta$ and He transitions are less evident, appearing as single-peaked lines, likely because of the lower signal-to-noise. We note that the H$\alpha$ line weakens by a factor $\sim$2 as the outburst evolves, from EW $\sim$4  to $\sim$2 \AA\ between May 4 and June 16, respectively (see Table \ref{tab:prop}).

The dip averaged spectra reveal the presence of broad absorption components in Balmer and \ion{He}{ii} that strengthen as the flux drops. Interestingly, no absorption is detected in \ion{He}{i}-5876 \AA, that keeps the same emission profile in both the dip and non-dip spectra. The Balmer and \ion{He}{ii} absorption minima reach 10 per cent below the continuum level (see Fig \ref{fig:spect}) and are all blue-shifted. 

In order to measure the centroid of the profiles we fitted a multi-Gaussian model to the  H$\alpha$, H$\beta$ and \ion{He}{ii} absorption components in the deepest (continuum normalised) dip spectrum of each night. The height, the FWHM, and the velocity offset of each Gaussian were set as free parameters. Instrumental broadening was taken into account in the fitting process. We note that the H$\alpha$ absorption is affected by a narrow emission component (red wing) that was masked.  The fits results are presented in Table \ref{tab:prop}. The FWHM of a given absorption also remains roughly consistent between the two epochs.  The terminal velocities of the absorption profiles range from $1579 \pm 180$ to $4178 \pm 222$~km~s$^{-1}$ (velocity of the blue edge at 10 per cent of the depth of the fitted Gaussian) for \ion{He}{ii}-4686 \AA\ and H$\beta$, respectively. On the other hand, the offset velocities obtained for the three lines are consistent with each other within $1.5 \sigma$. Thus, in a second step we tight the core velocities of the three Gaussians and performed a new fit. We derive velocity offsets of $840 \pm 101$ and $-766 \pm 59$ ~km~s$^{-1}$ for the first and second epoch, respectively.


\begin{table}
	\centering
	\caption{Properties of emission and absorption lines}
	\begin{tabular}{lccc} 
		\hline
		& FWHM & EW & Centroid velocity \\
		&(km~s$^{-1}$) & (\AA) & (km~s$^{-1}$) \\				

		\hline		
		Emission &&&\\
		\scriptsize{4 May, 12 and 16 June} &&&\\

		\hline
		
		\multirow{ 3}{*}{H$\alpha$}& $2824 \pm 92$ & $3.95 \pm 0.08$ & $-131 \pm 16$ \\		
		&$3059 \pm 142$& $3.53 \pm 0.13$ &$-313 \pm 32$\\		
		&$3084 \pm 281$& $2.17 \pm 0.16$  &$-12 \pm 63$\\
		
		\hline
		Absorption &&&\\
		\scriptsize{12 and 16 June}&&&\\
		\hline				
		
		\multirow{ 2}{*}{H$\alpha$}&$4031 \pm 399$ & $-11.0 \pm 0.6$  &$-908 \pm 227$ \\		
		&$3310 \pm 299$&$-10.0 \pm 0.4$&$-982 \pm 120$\\		
		&&& \\

		\multirow{ 2}{*}{H$\beta$}& $4380 \pm 444$  & $-5.8 \pm 0.4$ & $-690 \pm 177$\\		
		& $4585 \pm 344$ &$-4.4 \pm 0.4$&$-739 \pm 138$ \\		
		&&& \\

		\multirow{ 2}{*}{\ion{He}{ii}-4686 \AA}& $1790 \pm 446$ & $-2.3 \pm 0.4$  & $-841 \pm 158$\\		
		& $1733 \pm 198$ & $-3.3 \pm 0.3$ &$-725 \pm 75$\\		

		\hline
\end{tabular}
\label{tab:prop}
\end{table}

\begin{figure}
\includegraphics[scale=0.52]{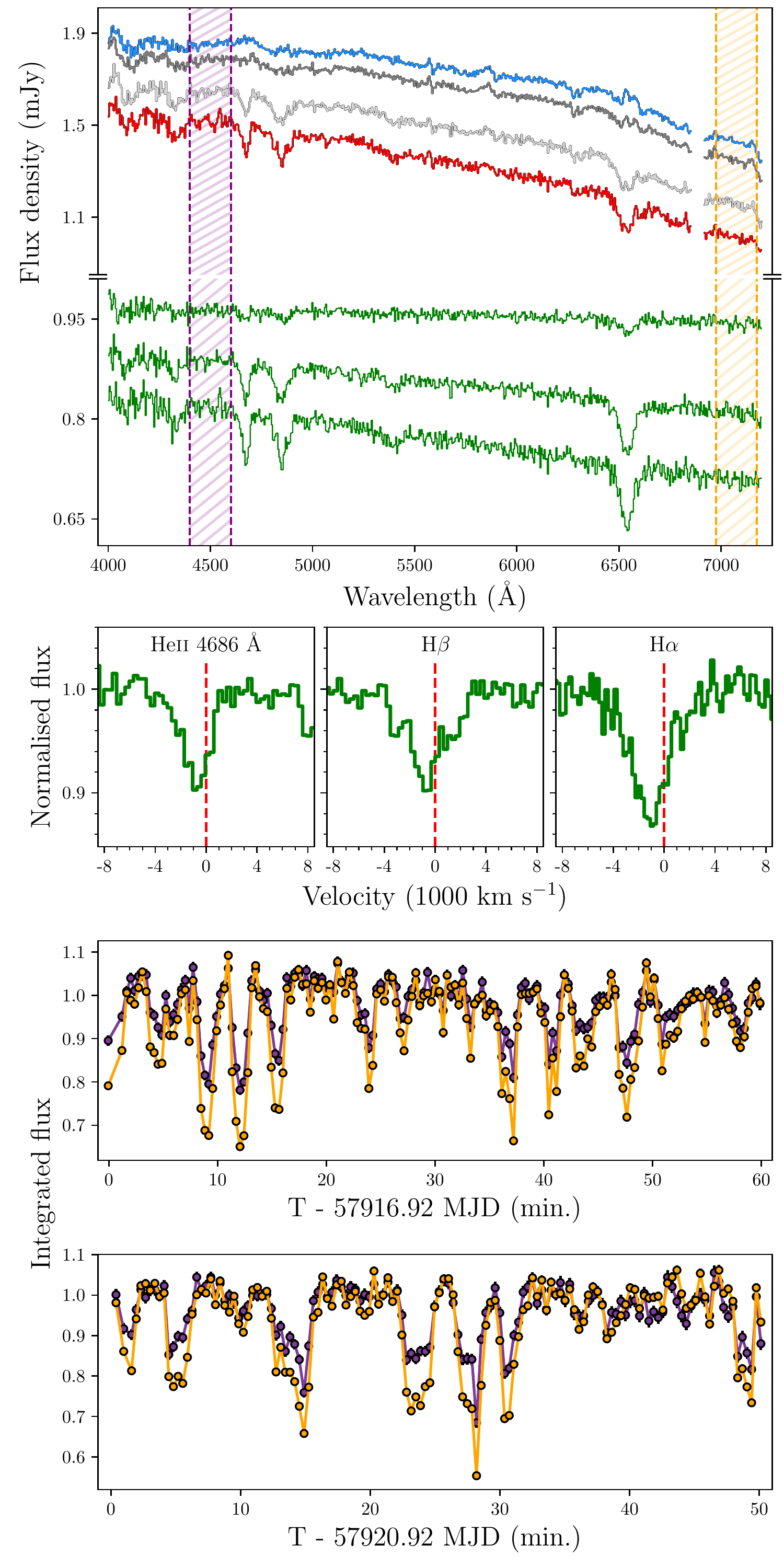}
\caption{Top panel: dip (red) and non-dip (blue) averaged spectra of the two GTC nights. The green spectra show the result of dividing the three dip spectra by the non-dip spectrum. Middle panel: H$\alpha$, H$\beta$, and \ion{He}{ii}-4686 \AA\ absorptions in the deepest (normalised) ratio-spectrum and represented in their corresponding line rest frames. Bottom panels: 12 June (top) and 16 June (bottom) light curves obtained from red and blue continuum windows in the spectra (orange and purple points, respectively), normalised to the non-dip level. The regions defining these two spectral windows are indicated as shaded areas in the top panel, following the same colour code as above.}
\label{fig:spectral_index}
\end{figure}

Finally, we have averaged the dip and the non-dip spectra of the two GTC nights and produced four final spectra, one for each defined class. The result is shown in Fig. \ref{fig:spectral_index} (top panel). The three average dip spectra were also divided by the average non-dip spectrum to produce three ratio spectra, that are plotted in green in the same figure. These reflect the evolution of the dip spectra relative to the non-dip spectrum, further enhancing spectral features caused by the dip while quenching those that remain unchanged. For example, the \ion{He}{i}-5876 \AA\ line vanishes in the ratio-spectra, while the blue-shifted absorptions become clearer in Balmer and \ion{He}{ii} (see Fig. \ref{fig:spectral_index}, middle panel). We also note a clear variation in the continuum slope of the ratio-spectra. This reveals a progressive change in the colour of the spectrum through the dip, indicating that the spectrum becomes more absorbed at red than at blue wavelengths.

To further examine this property, we have produced light curves from red and blue windows of the original spectra. We used \textsc{molly} software to integrate the flux density over two featureless spectral regions of width 200 \AA: the blue region is  centred at 5125 \AA\ while the red one at 7075 \AA\ (see Fig. \ref{fig:spectral_index}, top panel). The blue and red light curves are presented in the bottom panels of Fig. \ref{fig:spectral_index} for the two nights. In order to compare them, the light curves have been normalised to the non-dip level. We find that the dips obtained from the red part of the spectrum are $\sim$10 per cent deeper than those obtained from the blue part in both nights. In other words, the dips make the source bluer, contrary to what would be expected for the case of absorption by dust, which would redden the spectrum. This is entirely consistent with the results of \cite{Paice2019} based on multi-band high-time resolution photometry.

\section{Discussion} \label{sec:disc}

We have presented fast optical photometry and spectroscopy of Swift~J1357.2-0933 during its 2017 outburst. Our data show evidence for optical dips in the nine light curves obtained. The dips are analogous to the ones discovered during the 2011 outburst, they are quasi-periodic and repeat with a recurrence period that migrates to lower frequencies as the outburst declines. In \citetalias{Corral-Santana2013} these dips are interpreted as due to obscuration by a vertical disc structure (seen at high inclination) that propagates outward as the outburst evolves. The similarities between the 2011 and 2017 dips are not only in shape but also in time evolution (see Section \ref{subsec:opdibs}). The DRP follows the same cadence in both outbursts showing a similar frequency decay, which is remarkable for two phenomena occurring six years apart (see Fig. \ref{fig:phot}, bottom panel). This was also reported in \cite{Paice2019}.

We have resolved the dips using high time resolution optical spectroscopy in two epochs. The dip spectra show broad absorptions in the Balmer and the \ion{He}{ii} lines while no absorption is observed in \ion{He}{i}-5876 \AA\ (Fig. \ref{fig:spect}, lower panel). The absorptions become stronger as the flux drops within the dip and are conspicuous in spectra with a continuum level $\sim 3\sigma$ below the mean.

The absorption lines are blue-shifted, with centroid velocities of $\sim -800$~km~s$^{-1}$. Similar broad absorptions (sometimes with embedded emission) have been observed in other BHTs during outburst, e.g. GRO~J0422+32, XTE~J1118+480, and recently MAXI~J1807+132 \citep[][respectively]{Casares1995a,Dubus2001,Jimenez-Ibarra2019} but evolving at longer time scales \citep[e.g.][ detected them in a 3600~s exposure spectrum]{Jimenez-Ibarra2019}. In these systems, the absorptions are strongest at blue wavelengths, and show velocities consistent with the rest frame of the system, in contrast with what we seen in Swift~J1357.2-0933.

In the last decade, accretion disc winds have been established as a common feature in BHTs. In X-rays, \cite{Ponti2012} used a sample of high inclination BHTs to measure blue-shifted centroid velocities of $\sim-1000$~km~s$^{-1}$ in absorption features of highly ionised Fe among other species \citep[see also e.g.,][for similar studies]{Miller2006,Neilsen2009,DiazTrigo2016}. This is interpreted as the signature of hot equatorial winds.  In the optical, P-Cygni profiles have been reported in V404 Cygni, V4641 Sgr and MAXI J1820+070, demonstrating the presence of low-ionisation winds with terminal velocities in the range of $1000-3000$~km~s$^{-1}$ \citep{Munoz-Darias2016,MataSanchez2018,Munoz-Darias2018,Munoz-Darias2019}. Thus, the terminal velocity that we measure in Swift~J1357.2-0933 is consistent with the upper end of wind-type outflows observed in other BHTs.

While the observation of blue shifted Balmer absorptions in Swift~J1357.2-0933 can be also interpreted as an equatorial wind, the simultaneous presence of \ion{He}{ii} absorptions posses a challenge. \cite{Hoare1994} studied the \ion{He}{ii} transitions in non magnetic cataclysmic variables and found that some components of \ion{He}{ii}-4686 \AA\ originate in winds rather than in the disc. Furthermore, \cite{Drew1990} demonstrated that clumpy winds in O stars can enhance the strength of the recombination lines. In particular, \cite{Sundqvist2011} showed that H$\alpha$ reacts similarly to \ion{He}{ii}-4686 \AA\ to clumping. Our observations might fit into this scenario. On one hand, the \ion{He}{ii}-4686 \AA\ and H absorptions evolve quite similarly during the dip, reaching comparable depths within the same time scale. They also share dynamical properties such as equal (1.5$\sigma$) core velocities. All together, this might suggest a common origin for the absorptions.  However, the absorption is narrower in \ion{He}{ii}-4686 \AA\ than in Balmer (see Table \ref{tab:prop}). This might be explained if they arise from slightly different regions of the ejecta, while the \ion{He}{i}-5876 \AA\ emission would be formed in a different disc region since it is not affected by the dips. On the other hand, clumping would also explain the observed complex structure of the optical light curve. Deep but short lived dips with a sharp transition time-scale could be produced by dense inhomogeneities subtending a small angle over the line-of-sight. Within this context, the absorptions could be produced by optically thick and clumpy outflows.

If the winds are equatorial \citep[as seen in ][]{Ponti2012}, the outflowing  direction is close to the line-of-sight which is in agreement with the high (projected) velocities measured. In addition, this perspective favours the observation of wind inhomogeneities (clumps) which would be optically thick since we are looking across large columns of gas.

Assuming that the mean terminal velocity observed ($\sim 3000$~km~s$^{-1}$) corresponds to the escape velocity at the launching radius we infer $R_{l}$ $\sim$ 0.17 $R_{\odot}$. This assumption is expected to be roughly satisfied by some of the most popular wind launching mechanism \citep[e.g. thermal winds,][]{Begelman1983}. On the other hand, assuming that the DRP corresponds to the Keplerian frequency of a given disc annulus, the values measured on 12 and 16 June would correspond to Keplerian radii at $R_{P} \sim 0.16-0.18$ $R_{\odot}$. The remarkable agreement reached from very different observables suggest that winds and dips are produced at the same radius. We propose two explanations for the dipping phenomenology in Swift~J1357.2-0933, within the equatorial wind scheme. (i) The disc is warped by some mechanism that generates a vertical structure \citepalias{Corral-Santana2013}. Viewed at high inclination, this structure hides the inner disc, producing the dips. The gas in the vertical extend is exposed to irradiation and is expelled radially along the equatorial plane. (ii) Winds are triggered by some mechanism at a given radius, where the upper disc layers are expelled. The action of the winds therefore shapes the disc at a particular radius, producing dips in a high inclination perspective.

Our study reveals a colour-dependence behaviour of the dips. The red wavelengths are more obscured during the dip than the blue ones. This is at odds with expectations from standard dust scattering, since there is no standard dust extinction law that can explain this behaviour. In the extreme case, grey extinction (i.e. optically thick material at all the observed wavelengths) will produce a flat continuum with no change in the slope of the ratio-spectra. The colour evolution of the dips is also evident after comparing synthetic light curves obtained from two continuum apertures in the red and blue part of the spectra (see Section \ref{sec:spec}). During the dip, the red lightcurve always reaches levels $\sim 10$ per cent lower than the blue lightcurve. Qualitatively, the deeper the dip, the larger the contrast with its blue counterpart.  \cite{Paice2019} put forward an scenario to explain the observed bluing of the source during the dip. The occultation of the lower (and brighter) part of the jet by the dipping structure would prevent the red component from contributing to the total spectral energy distribution, producing bluer spectra.  In this framework, if we assume that only the red component is absorbed, the subtraction of the dip from the non-dip spectrum would yield the absorbed component, multiplied by a function of the extinction law. 
By doing this, we obtained four residual spectra adopting four common Milky Way dust extinction models. We used the fuctions contained in the package \textsc{astropy-dust extinction} \citep[CCM89, O94, F99, and F04 from][respectively]{Cardelli1989,ODonnell1994,Fitzpatrick1999,Fitzpatrick2004}. We used the standard value R$_{\mathrm{V}}$=3.1 and considered A$_{\mathrm{V}}$=0.3~mag (average magnitude drop during dips; see Fig. \ref{fig:phot}). The continuum of the residual spectra are well described by a power--law ($F_\nu \propto \nu^ {\alpha}$) with $-1.6<\alpha<-1.4$. All considered, this result is not far from predictions for jet emission in the optically thin regime \citep[$-1 < \alpha < -0.4$;][]{Blandford1979}, and is consistent with the spectral index measured in this system during quiescence by \citet{Shahbaz2013}, using near-IR-optical data ($\alpha =- 1.4 \pm 0.1$).

However, it is important to bear in mind that the above scenario assumes that the optical dips observed in the continuum are created by dust scattering, while we also observe absorption lines indicating the presence of outflowing, ionised gas. To start with, it is not clear how dust can be present in the hot environment surrounding BHTs. \footnote{However, we note that dust has been recently found in a quintuplet of Wolf-Rayet stars \citep{Najarro2017}} Furthermore, our observations show that the behaviour of the continuum is clearly correlated with that of the absorption lines. A simpler alternative scenario would be that the continuum is absorbed by the same ionised gas producing the absorption lines via bound-free interactions, similar to those present in stellar atmospheres at temperatures low enough to allow the formation of recombination lines of hydrogen and helium. Interestingly, the effective cross-section of this interaction (for hydrogen) in the optical range increases with wavelength \citep[e.g.][]{Gray2005}, which could naturally explain the blue colour observed during the dips. Nevertheless, we note that a more detailed radiative modelling beyond the scope of this work would be necessary to support this latter scenario.

It is worth mentioning that, besides other examples of variability observed in the optical spectra of BHT -- of which Swift~J1357.2-0933 displays one of the most dramatic cases --, growing evidence for a qualitatively similar phenomenology is being found in super-massive black holes (including the so-called \textit{changing look quasars}). For instance, \citet{Vivek2018} observed broad transient absorptions in optical spectra of the quasar SDSS J133356.02+001229, varying on time-scales from a few days to years. This was interpreted as being produced by clouds crossing the line of sight in a co-rotating clumpy wind. Despite the fact that analogies between observables in stellar-mass and super-massive black holes need to be taken with caution, these observations might suggest that similar clumpy winds are common across a wide range of black hole masses.

Finally, during the review process of this work a paper presenting consistent phenomenology with that reported here was published by Charles et al. (2019; in press).

\section{Conclusions} \label{sec:conc}
 We have presented photometric and spectroscopic observations of Swift~J1357.2-0933 during the decay of its 2017 outburst. In agreement with previous studies, we detect the presence of optical dips with a recurrence time that gradually increases throughout the outburst. The dips  have a blue colour and its evolution resembles that observed in the 2011 event \citepalias[][\citealt{Paice2019}]{Corral-Santana2013}.
Our dip-resolved spectroscopic study indicates the existence of an outflow associated with the appearance of optical dips during the 2017 outburst of the black hole transient Swift~J1357.2-0933. We interpret the optical dips as produced by a clumpy and dense equatorial wind. Within this scenario, the detection of dips is strongly dependent on orbital inclination. Therefore, this clumpy equatorial wind might not be a peculiar characteristic of this system but a common phenomenon of accreting stellar-mass black holes in outburst.

\section{Acknowledgements}
We acknowledge support by the Spanish MINECO under grant AYA2017-83216-P. TMD acknowledge support via Ram\'{o}n y Cajal Fellowships RYC-2015-18148. Based on observations made with the GTC telescope, in the Spanish Observatorio del Roque de los Muchachos of the Instituto de Astrof\'{i}sica de Canarias, under Director's Discretionary Time. We are thankful to Miriam Garcia for useful discussion.




\bibliographystyle{mnras}
\bibliography{biblio.bib} 



\bsp	
\label{lastpage}
\end{document}